\documentclass[12pt]{article}

\usepackage{epsfig}

\newcommand{\csection}[1]{\begin{center}{\Large\bf #1}\end{center}}

\newcommand{\bold}[1]{\mbox{\bf #1}}
\newcommand{\grbold}[1]{\mbox{\boldmath $#1$}}

\begin{document}

\title{\bf Potential Momentum, Gauge Theory, and Electromagnetism
in Introductory Physics}
\author{David J. Raymond \\
Physics Department \\
New Mexico Tech \\
Socorro, NM 87801}
\maketitle

\csection{Abstract}

\begin{quotation}
If potential energy is the timelike component of a four-vector, then
there must be a corresponding spacelike part which would logically be
called the \emph{potential momentum}.  The \emph{potential
four-momentum} consisting of the potential momentum and the potential
energy taken together, is just the gauge field of the associated force
times the charge associated with that force.  The canonical momentum
is the sum of the ordinary and potential momenta.

Refraction of matter waves by a discontinuity in a gauge field can be
used to explore the effects of gauge fields at an elementary level.
Using this tool it is possible to show how the Lorentz force law of
electromagnetism follows from gauge theory.  The resulting arguments
are accessible to students at the level of the introductory
calculus-based physics course and tie together classical and quantum
mechanics, relativity, gauge theory, and electromagnetism.  The
resulting economy of presentation makes it easier to include modern
physics in the one year course normally available for teaching
introductory physics.
\end{quotation}

\section{Introduction}

Many physicists believe that it is important to introduce more modern
physics to students in the introductory college physics course.  (See,
for instance, the summary of the IUPP project by Coleman et
al. \cite{coleman}.)  Our experience indicates that retaining the
conventional course structure and tacking modern physics on at the end
almost guarantees failure in this endeavor, due to the time
constraints of the typical one year course.  Completely restructuring
the course has been tried (see, for instance, Amato \cite{amato},
Mills \cite{mills}, Moore \cite{moorea}, Weinberg \cite{weinberg83})
and some form of this approach may be the most effective way to
accomplish the desired goal.

Raymond and Blyth \cite{raymondblyth} briefly outlined efforts to
develop an introductory college physics course with a radically modern
perspective.  The first semester of this two semester course begins
with optics and waves, proceeds from there to relativistic kinematics,
and then introduces the basic ideas of quantum mechanics.  It finishes
with the development of classical mechanics as the geometrical optics
limit of quantum mechanics.  The second semester builds on the results
of the first semester and covers more mechanics, electromagnetism, a
survey of the standard model, and statistical physics.

The structure of this course was inspired by the Nobel Prize address
of Louis de Broglie \cite{debroglie}.  The logic which led de Broglie
to the relation between momentum and wavenumber makes use of a
beautiful application of the principle of relativity to obtain the
relativistic dispersion relation for waves representing massive
particles.  Furthermore, de Broglie largely anticipated the
development of the Schr\"{o}dinger equation via an analogy between the
propagation of light through a medium of varying refractive index and
the propagation of matter waves in a region of spatially variable
potential energy.

De Broglie's approach to matter waves subject to a potential suggests
a way of getting these ideas across in elementary form to beginning
physics students.  We believe that the use of the Schr\"{o}dinger
equation per se is inappropriate at the level of an introductory
physics course.  However, much of the physics of this equation can be
extracted from its dispersion relation in the case where the potential
energy is piecewise constant.

The dispersion relation for free matter waves was obtained by de
Broglie from little more than the notion of relativistic invariance.
Relativity also helps in the development of a modified dispersion
relation for matter waves subject to external influences, since the
modified as well as the original relation must satisfy relativistic
invariance.  There turn out to be only a few possible invariant ways
to make this modification --- gauge theory results from one of these.
Pursuit of this choice leads inevitably to the Lorentz force law of
electromagnetism.  Use of Coulomb's law and variations on the
relativistic invariance arguments of Purcell \cite{purcell} and
Schwartz \cite{schwartz} lead to most of the rest of electromagnetism.

The key arguments in this path involve the derivation and use of
variations on Snell's law to show how wave packets change their
velocity when they cross a discontinuity in a gauge field.  These
arguments invoke the same physics that appears in the application of
Hamilton's principle to the generation of the equations of mechanics,
albeit on a much more elementary level.  This connection becomes
evident when it is realized that Snell's law contains the same physics
as Fermat's principle, which in turn is an application of the
principle of least action.

The above comments explain why we initially study waves and relativity
in the introductory course.  Mastery of these topics allows even
beginning students to obtain at least some insight into the most
subtle and profound ideas of physics, insight which can only be
acquired via the traditional path after many years of hard work.  In
addition, it points to an alternative approach to electromagnetism at
the introductory level which may actually be easier for beginning
students to understand than the traditional approach and is
undoubtedly more economical of student time and effort.

\section{Potential Momentum}

De Broglie treated particles in a potential by developing a formula
for the index of refraction $n$ such that wave packets of light would
act like such particles.  The problem of computing the evolution of
matter waves was then reduced to the problem of computing the
propagation of light through a medium with variable refractive index.
Pursuit of this program leads to the Schr\"{o}dinger equation.

A great deal can be learned about the propagation of light through a
spatially variable medium by examining the case of piece-wise constant
index of refraction.  In a region in which $n ( \bold{k} )$ is
independent of position, wave packets of light move according to the
group velocity computed from the dispersion relation $\omega = c |
\bold{k} | /n$, where $\omega$ is the angular frequency of the light,
$c$ is the speed of light in a vacuum, and $\bold{k}$ is its wave
vector.  A wave packet crossing a discontinuity in $n$ is treatable
as a problem in refraction.

De Broglie concentrated on the non-relativistic limit in his
discussion of the interaction of waves with a potential.  However, in
some ways the relativistic case is simpler, because any modification
to the free particle dispersion relation for relativistic matter waves
must also be relativistically invariant.  The dispersion relation for
free relativistic particles of mass $m$ is
\begin{equation}
	\omega^2 - k^2 c^2 = m^2 c^4 / \hbar^2
\end{equation}
where $\hbar$ is Planck's constant divided by $2 \pi$.  We write this
in the notation of the introductory course as
\begin{equation}
	\underline{k} \cdot \underline{k} = -m^2 c^2 / \hbar^2
\label{freeparticle}
\end{equation}
where $\underline{k} = ( \bold{k} , \omega /c )$ is the \emph{wave
four-vector} with the wave vector $\bold{k}$ being the spacelike part
and $\omega /c$ the timelike part of $\underline{k}$.  (We find it
advantageous to use Weinberg's choice of metric \cite{weinberg95},
$\underline{k} \cdot \underline{k} = \bold{k} \cdot \bold{k} -
\omega^2 /c^2$, because the four-vector dot product becomes a simple
extension of the ordinary dot product with one additional term.)

One can imagine only a few ways to modify equation
(\ref{freeparticle}) so as to maintain relativistic invariance:
\begin{itemize}
\item
One could write
\begin{equation}
	\underline{k} \cdot \underline{k} = -(m - A)^2 c^2 / \hbar^2
\label{scalartheory}
\end{equation}
where $A$ is a relativistic scalar.
\item
An alternative would be
\begin{equation}
	( \underline{k} - \underline{A} ) \cdot
	( \underline{k} - \underline{B} ) = -m^2 c^2 / \hbar^2
\end{equation}
where $\underline{A}$ and $\underline{B}$ are four-vectors.  Weak,
strong, and electromagnetic gauge theories correspond to the case
$\underline{A} = \underline{B}$.
\item
The final apparent possibility is
\begin{equation}
	\underline{k} \cdot \underline{N} \cdot \underline{k} = -m^2
	c^2 / \hbar^2
\label{tensortheory}
\end{equation}
where $\underline{N}$ is a tensor which becomes the identity tensor of
spacetime in the free particle case.
\end{itemize}

In the introductory course we state that
\begin{equation}
	( \underline{k} - \underline{T} ) \cdot
	( \underline{k} - \underline{T} ) = -m^2 c^2 / \hbar^2
\end{equation}
is the correct choice for all natural forces except gravity.  However,
we also explore the consequences of a scalar potential field.  (The
tensor potential is too complex to consider in an introductory
course!)

The four-vector gauge field $\underline{T}$ has the form
$\underline{T} = ( \bold{T} , S/c )$ so that the dispersion relation
for matter waves under the influence of a constant gauge field expands
to
\begin{equation}
	(\omega - S)^2 - | \bold{k} - \bold{T} |^2 c^2 = m^2 c^4 /
	\hbar^2 .
\end{equation}
Multiplying this equation by $\hbar^2$ leads to the relativistic
energy-momentum relationship for particles interacting with a constant
gauge field,
\begin{equation}
	(E - U)^2 - | \grbold{\Pi} - \bold{Q} |^2 c^2 = m^2 c^4 ,
\label{massenergymom}
\end{equation}
where $\underline{Q} \equiv \hbar \underline{T} = (\bold{Q} , U/c)$.
$E = \hbar \omega$ is the total energy and $\grbold{\Pi} = \hbar
\bold{k}$ is the canonical momentum \cite{goldstein}.  However, we
refer to it as the \emph{total momentum} in the introductory course in
analogy with the total energy.  The quantity $U$ is just the potential
energy.

We call the new quantity $\bold{Q}$ the \emph{potential momentum}.
This terminology arose in a natural way during the first iteration of
the introductory course.  We had discussed the fact that momentum and
kinetic energy together form a four-vector.  I then pointed out that
potential energy is a quantity much like kinetic energy and asked the
class what this implied.  A particularly bright freshman replied
``\ldots groan, potential momentum!''.

\section{Group Velocity and Refraction of Matter Waves}

In the introductory class it is necessary at this point to teach the
students about partial derivatives, since they typically have not yet
covered this concept in their calculus course.  However, this is a
relatively easy idea to get across.  It is also necessary to extend
the idea of group velocity (about which they have learned earlier in
the section on optics) to more than one dimension.  Thus, for
instance, in two dimensions we have $\omega = \omega (k_x , k_y )$ and
the group velocity is
\begin{equation}
	\bold{u}_g = \left( \frac{\partial \omega}{\partial k_x} ,
		\frac{\partial \omega}{\partial k_y} \right) .
\label{groupvel}
\end{equation}
At this stage we find it best to state the above result without proof
as a plausible extension of the one-dimensional result.

\begin{figure}
\begin{center}
\psfig{figure=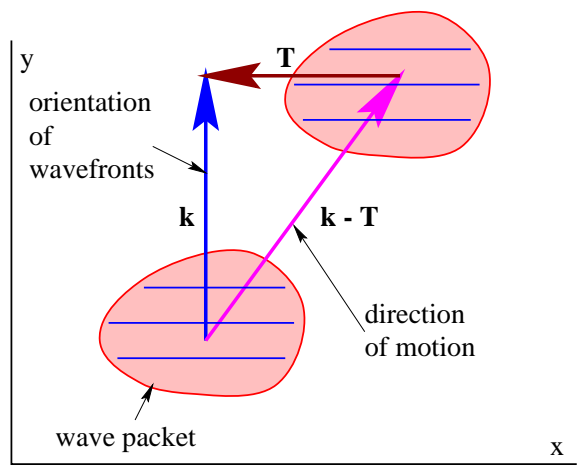,width=3in}
\end{center}
\caption{Orientation and motion of a quantum mechanical wave packet
in two dimensions.  The vector $\bold{T}$ is perpendicular to the
wave vector $\bold{k}$ in this case.}
\label{wavepacket3d}
\end{figure}

Given equation (\ref{groupvel}) it is easy to show that the group
velocity of a quantum mechanical wave is just
\begin{equation}
	\bold{u}_g = \frac{(\bold{k} - \bold{T})c^2}{\omega - S} =
	\frac{( \grbold{\Pi} - \bold{Q})c^2}{E - U} .
\label{relgv}
\end{equation}
This reduces to the usual expression for the group velocity of
relativistic matter waves when $\bold{Q}$ and $U$ are zero.

An example of the motion of a wave packet is illustrated in figure
\ref{wavepacket3d}, which shows the student that wave packets don't
necessarily move in the direction of the wave vector when $\bold{T}
\ne 0$.

Eliminating $E - U$ and then $| \grbold{\Pi} - \bold{Q}|$ between
equations (\ref{massenergymom}) and (\ref{relgv}) yields
generalizations of the usual formulas for the relativistic momentum
and energy,
\begin{equation}
	\grbold{\Pi} = \bold{Q} + m \bold{u}_g \gamma
\end{equation}
\begin{equation}
	E = U + mc^2 \gamma
\end{equation}
where $\gamma = (1 - u_g^2 /c^2 )^{-1/2}$.  We call the ordinary
relativistic momentum $\bold{p} \equiv m \bold{u}_g \gamma$ the
\emph{kinetic momentum} just as $m u_g^2 /2 = E - U = K$ is the
kinetic energy.

\begin{figure}
\begin{center}
\psfig{figure=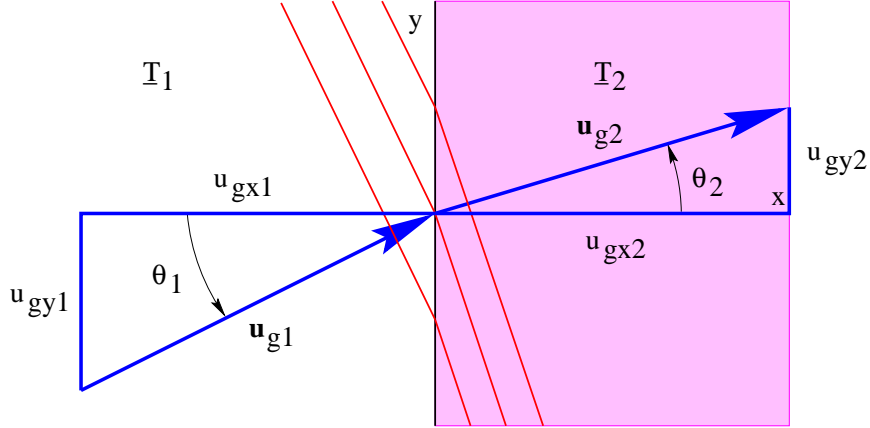,width=4.5in}
\end{center}
\caption{Refraction by a discontinuity in a gauge field
$\underline{T}$ at $x = 0$.}
\label{grefract}
\end{figure}

Figure \ref{grefract} shows refraction by a discontinuity in a gauge
field.  The frequency, $\omega$, and the component of the wave vector
parallel to the interface, $k_y$, are constant across the interface as
a result of phase continuity.  This implies that $E$ and $\Pi_y$ are
continuous as well.  In the non-relativistic limit $\gamma \approx 1 +
u_g^2 / (2c^2 )$, and these conditions reduce to the continuity of $U
+ m u_g^2 /2$ and $Q_y + m u_{gy}$ across the interface.

\section{Gauge Forces, Non-Relativistic Limit}

\begin{figure}
\begin{center}
\psfig{figure=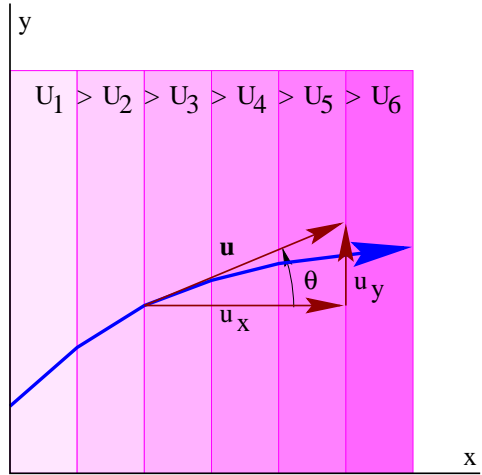,width=2.5in}
\end{center}
\caption{Trajectory of a wave packet through a variable scalar gauge
field in which $U = \hbar S$ decreases to the right.}
\label{gscalar}
\end{figure}

The above expressions are sufficient to infer the accelerations of
a particle in the non-relativistic geometrical optics limit, and
therefore the forces acting on it.  Let us first set $\bold{Q} = 0$
and approximate continuous variability in $U$ along the $x$ axis by a
sequence of slabs of constant $U$ as shown in figure \ref{gscalar}.
In this case $U + m u_g^2 /2$ and $u_{gy}$ are constant.  Thus, the
$x$ and $y$ components of the force are
\begin{eqnarray}
	F_x & = & m \frac{d u_{gx}}{dt} = m \frac{d u_{gx}}{dx}
	\frac{dx}{dt} = m \frac{d u_{gx}}{dx} u_{gx} \nonumber \\
	& = & \frac{m}{2} \frac{d u_{gx}^2}{dx} =
	\frac{m}{2} \frac{d u_g^2}{dx} = - \frac{dU}{dx}
\end{eqnarray}
and
\begin{equation}
	F_y = m \frac{d u_{gy}}{dt} = 0 .
\end{equation}
In this case the force is the familiar negative gradient of the
potential energy.

\begin{figure}
\begin{center}
\psfig{figure=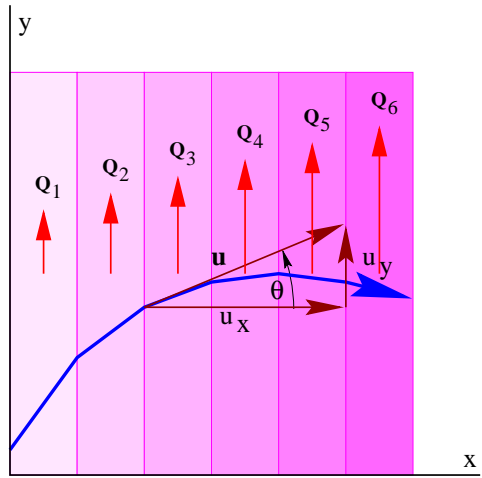,width=2.5in}
\end{center}
\caption{Trajectory of a wave packet through a variable vector gauge
field in which $\bold{Q} = \hbar \bold{T}$ changes to the right as
shown.}
\label{gvector}
\end{figure}

Next we set $U = 0$ and assume that $\bold{Q} = (0,
Q , 0)$, with $Q$ varying as shown in figure \ref{gvector}.  Since
$Q + m u_{gy}$ is constant in the non-relativistic limit, we see that
\begin{equation}
	F_y = m \frac{d u_{gy}}{dt} = m \frac{d u_{gy}}{dx} u_{gx} =
	-u_{gx} \frac{dQ}{dx} .
\end{equation}
However, $mu_g^2 /2 = m ( u_{gx}^2 + u_{gy}^2 )/2$ is also constant in
this case, from which we infer that
\begin{equation}
	F_x = m \frac{d u_{gx}}{dt} = - m \frac{u_{gy}}{u_{gx}}
	\frac{d u_{gy}}{dt} = -m u_{gy} \frac{d u_{gy}}{dx} = u_{gy}
	\frac{dQ}{dx} .
\end{equation}
Putting this together, we find
\begin{equation}
	\bold{F} = \left( u_{gy} \frac{dQ}{dx} , - u_{gx}
	\frac{dQ}{dx} , 0 \right) .
\label{gvforce1}
\end{equation}

This is a special case of the more general result
\begin{equation}
	\bold{F} = \bold{u}_g \times \bold{P}
\end{equation}
where
\begin{equation}
	\bold{P} = \left( \frac{\partial Q_z}{\partial y} -
		\frac{\partial Q_y}{\partial z} ,
		\frac{\partial Q_x}{\partial z} -
		\frac{\partial Q_z}{\partial x} ,
		\frac{\partial Q_y}{\partial x} -
		\frac{\partial Q_x}{\partial y}
	\right) .
\end{equation}
The general case is not proved in the introductory class, but is
presented as a plausible generalization of equation (\ref{gvforce1}).

\begin{figure}
\begin{center}
\psfig{figure=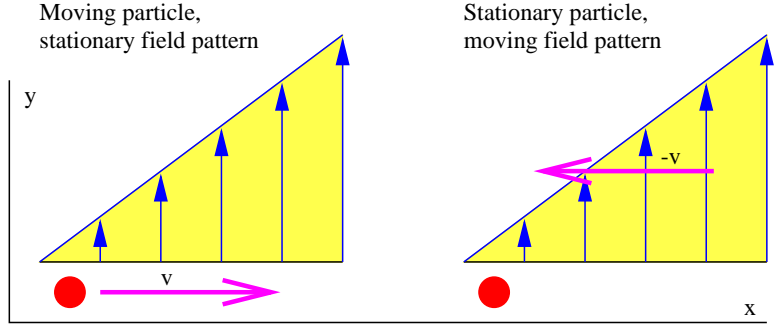,width=4in}
\end{center}
\caption{A moving particle and a stationary pattern of potential
momentum $\bold{Q} (x)$ must be equivalent to a stationary particle
and a moving pattern of potential momentum according to the principle
of relativity.}
\label{gtrans}
\end{figure}

So far we have assumed that the gauge fields affecting particles are
constant in time.  An interesting effect occurs when a vector gauge
field varies with time.  Figure \ref{gtrans} shows how we attack this
problem.  According to the principle of relativity, the case of a
particle moving in the $+x$ direction through a region in which
$\bold{Q}$ is steady with time but increases in magnitude with $x$ is
equivalent to the case of a \emph{stationary} particle in which
$\bold{Q}$ at the particle increases with time.  As figure
\ref{gtrans} shows, this is because in the latter situation the whole
pattern of $\bold{Q}$ shifts to the left with time, resulting in the
particle being exposed to larger and larger values of $\bold{Q}$.  (We
emphasize to our classes that it is important to think of the field
\emph{pattern} as shifting, not the field itself.  Fields don't move,
they just have space and time variations which look different in
different reference frames.)  The time rate of change of $\bold{Q}$ at
the position of the particle in the right panel of figure
\ref{gtrans} is
\begin{equation}
	\frac{\partial \bold{Q}}{\partial t} = \frac{\partial
	Q}{\partial x} \frac{dx}{dt} = v \frac{\partial
	\bold{Q}}{\partial x} .
\label{advection}
\end{equation}

Equation (\ref{gvforce1}) applies to the left panel of figure
\ref{gtrans} with $u_{gx} = v$, $u_{gy} = 0$ and $\bold{Q} = (0,Q,0)$.
The resulting force on the object is $\bold{F} = -v ( \partial
\bold{Q} / \partial x)$.  However, by the principle of relativity, the
force on the object in the right panel should be the same.  (We are
assuming low velocity transformations, so the issue of $\bold{F}$
possibly changing as a result of the transformation doesn't arise
here.)  We therefore infer that the force on a stationary particle
with time-varying $\bold{Q}$ is
\begin{equation}
	\bold{F} = - \frac{\partial \bold{Q}}{\partial t} .
\end{equation}

Putting all these effects together, the complete gauge force is thus
written
\begin{equation}
	\bold{F} = - \frac{\partial \bold{Q}}{\partial t} -
	\left( \frac{\partial U}{\partial x} ,
	\frac{\partial U}{\partial y} ,
	\frac{\partial U}{\partial z} \right) +
	\bold{u}_g \times \bold{P} ,
\end{equation}
in the non-relativistic case, which is recognizable as being only a
step away from the Lorentz force law for electromagnetism.  If the
charge on the particle is $q$, then the scalar potential is $\phi =
U/q$, the vector potential is $\bold{A} = \bold{Q}/q$, the electric
field $\bold{E}$ is the first two terms on the right side of the above
equation divided by $q$, and the magnetic field is $\bold{B} =
\bold{P}/q$.  Thus, using only elementary arguments, the Lorentz force
is shown to be a consequence of the gauge theory assumption, which in
turn arises from the assumption of a potential momentum.

\section{Electromagnetism}

Purcell \cite{purcell} and Schwartz \cite{schwartz} infer the
character of electromagnetic fields from moving charges by performing
a Lorentz transformation on configurations of stationary charges.
This trick becomes considerably easier when attention is focused on
the four-potential $\underline{a} = ( \bold{A} , \phi /c )$ rather
than on the electric and magnetic fields, because the transformation
properties of the four-potential are much simpler.

\begin{figure}
\begin{center}
\psfig{figure=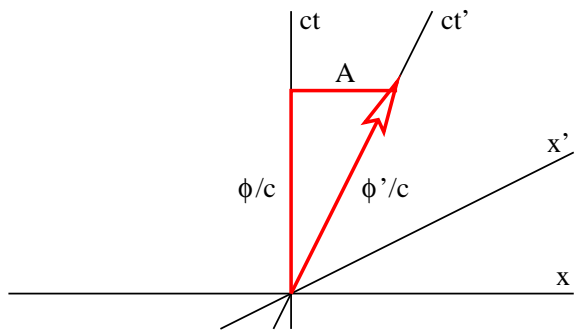,width=3in}
\end{center}
\caption{Finding the space and time components of the four-potential
which points in the time-like direction in the primed reference frame.
The $ct'$ axis is the world line of the charge which generates the
four-potential.  The primed frame moves to the right at speed $v$.}
\label{boosta}
\end{figure}

Figure \ref{boosta} illustrates the idea.  The four-potential in this
figure points along the $ct'$ axis and has invariant length $\phi'
/c$.  Since the primed reference frame moves to the right at speed
$v$, the components of the four-potential in the unprimed frame are
$\underline{a} = [ ( \beta \gamma \phi' /c ) \bold{i} , \gamma \phi'
/c ]$ where $\beta = v/c$, $\gamma = (1 - \beta^2 )^{-1/2}$, and
$\bold{i}$ is the unit vector in the $x$ direction.

\begin{figure}
\begin{center}
\psfig{figure=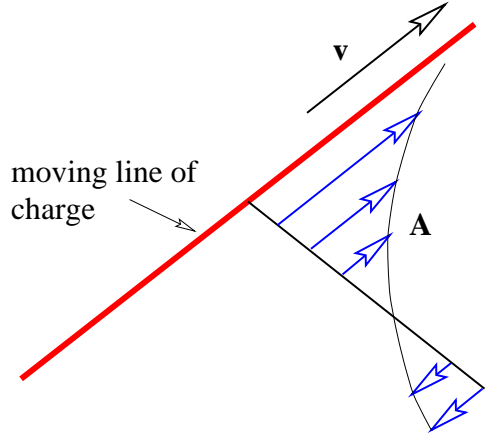,width=2.5in}
\end{center}
\caption{Vector potential from a moving line of charge.  The
distribution of vector potential around the line is cylindrically
symmetric.}
\label{linea}
\end{figure}

Let us consider a line of charge along the $x$ axis moving with the
primed reference frame.  In this reference frame we define the linear
charge density to be $\lambda'$ and find by the usual Gauss's law
techniques plus a simple integration that the scalar potential in the
primed frame is
\begin{equation}
	\phi' = -\frac{\lambda'}{2 \pi \epsilon_0} \ln r
\end{equation}
where $r$ is the distance from the $x$ axis.  We immediately infer
from figure \ref{boosta} that the vector potential in the unprimed
frame is
\begin{equation}
	\bold{A} = - \frac{\beta \gamma \lambda' \bold{i}}{2 \pi
	\epsilon_0 c} \ln r = - \frac{v \lambda \bold{i}}{2 \pi
	\epsilon_0 c^2} \ln r ,
\end{equation}
where $\lambda = \gamma \lambda'$ is the Lorentz-contracted charge
density in this frame.  The vector potential is illustrated in figure
\ref{linea} and is easily shown to yield the classical result for the
magnetic field due to a current $i = v \lambda$:
\begin{equation}
	| \bold{B} | = \frac{v \lambda}{2 \pi \epsilon_0 c^2 r} =
	\frac{\mu_0 i}{2 \pi r} .
\end{equation}
We have used $\epsilon_0 \mu_0 = c^{-2}$ where $\mu_0$ and
$\epsilon_0$ are the permeability and permittivity of free space.

\section{Discussion}

Some may question the heavy dependence on arguments based on
relativity in the present approach.  However, our experience has been
that relativity is avidly absorbed and understood even by average
beginning students when presented in terms of spacetime triangles and
a ``spacetime Pythagorean theorem'' rather than in terms of the rather
more abstract Lorentz transformations.  Four-vectors don't seem to
present any particular problems and arguments based on relativistic
invariance seem to resonate with the students.  In this respect our
experiences are similar to Moore's \cite{mooreb}.

Others may point out that our treatment of gauge theory leaves out the
connection between gauge invariance and the form of the Lagrangian.
We confess to not having figured out how to present this result in a
way which makes sense to the average college freshman.  However, we do
point out (in a problem set) that the form of the four-potential
leading to particular electric and magnetic fields is not unique.
Furthermore, our approach makes non-classical phenomena such as the
Aharonov-Bohm effect relatively easy for beginning students to
understand.

There are at least three major advantages to the presentation proposed
here:
\begin{enumerate}
\item
Certain aspects of gauge theory are touched upon in a way that makes
sense to a beginning student.  This is desirable in that the nearly
universal role of gauge theory in representing the forces of nature is
otherwise hard to describe in a non-trivial way at an introductory
level.
\item
Our route is arguably a more compact and insightful way to approach
electromagnetism than the normal
presentation.  This is an important consideration given the limited
time typically available to the introductory course.
\item
The mathematics used here is arguably simpler than is seen in typical
presentations of the mechanics of conservative forces and
electromagnetism.  In particular, though partial derivatives are
introduced, the use of the line integral, a particularly puzzling
concept for beginning students, is completely avoided.
\end{enumerate}
Our approach thus fosters the goal of presenting the most profound
ideas of physics in a manner that is as accessible as possible to
beginning students.

{\em Acknowledgments}: Particular thanks go to Robert Mills of Ohio
State University for pointing out a serious error in an earlier
version of this paper.

\end{document}